\documentclass[conference,letterpaper]{IEEEtran}
\usepackage[T1]{fontenc}
\usepackage[utf8]{inputenc}
\usepackage{cite}
\usepackage{amsmath,amssymb,amsthm}
\usepackage{algorithm}
\usepackage[noend]{algpseudocode}
\usepackage{booktabs}
\usepackage{graphicx}
\usepackage[table]{xcolor}
\definecolor{rowshade}{gray}{0.95}
\algrenewcommand{\algorithmiccomment}[1]{\unskip\nobreak\hfill\hspace{.3em}\mbox{\small\textcolor[gray]{.4}{\text{//} #1}}}
\makeatletter
\providecommand{\insert@pcolumn}{\insert@column}
\newcommand{\compactalgorithmic}{%
  \def\ALG@doentity{%
    \edef\ALG@thisblock{\csname ALG@currentblock@\theALG@nested\endcsname}%
    \expandafter\ifx\csname ALG@b@\ALG@L @\ALG@thisentity @\ALG@thisblock\endcsname\relax
      \def\ALG@thisblock{0}%
    \fi
    \ALG@getentitytext
    \ifnum\ALG@thisblock=0\else\ALG@doend\fi
    \ifx\ALG@text\ALG@x@notext
    \else
      \item\noindent\hskip\ALG@tlm
    \fi
    \expandafter\ifnum0=\csname ALG@b@\ALG@L @\ALG@thisentity @\ALG@thisblock\endcsname
    \else\ALG@dobegin\fi
    \def\ALG@entitiecommand{\ALG@displayentity}%
  }%
}
\makeatother
\newcommand{\rangeval}[1]{{\fontsize{7.5}{8.5}\selectfont\color[gray]{.35}(#1)}}
\usepackage[hidelinks]{hyperref}

\newtheorem{theorem}{Theorem}
\newtheorem{lemma}{Lemma}

\theoremstyle{definition}

\title{Optimal-Time Independent Subset Sampling over Acyclic Joins}
\author{Xuyang Liu$^{\dagger,\ddagger}$, Zijian Zhang$^{\dagger,*}$, Zhen Li$^\dagger$\\
	$^\dagger$School of Cyberspace Science and Technology, Beijing Institute of Technology, China\\
	$^\ddagger$School of Computer Science, The University of Auckland, New Zealand\\
	Email: \{liuxuyang, zhangzijian, zhen.li\}@bit.edu.cn\\
	$^*$Corresponding author
}

\begin{document}

\maketitle
\begin{abstract}
Join results can greatly exceed their inputs even when applications need only a small random subset.
We study sampling that includes each result independently with a prescribed probability given by the product of its input-tuple weights (PRODUCT).
For fixed full acyclic joins with static weights, we achieve input-linear preprocessing and index space together with output-linear expected sampling work, eliminating logarithmic factors from previous bounds.
A shared join-tree index supports weighted draws over the entire join and locates high-probability results.
Direct Bernoulli trials handle high-probability results, allowing Poisson thinning to generate the remaining independent inclusion decisions with bounded overhead.
Our exact sampler uses expected $O(N)$ preprocessing, $O(N)$ index space, and expected $O(1+m\mu)$ time per call including output construction.
Here $N$ is the input size, $m$ counts relation occurrences, and $\mu$ is the expected sample size.
The guarantees hold on a real random-access machine with unit-cost exact arithmetic and logarithms, continuous uniform randomness, and expected-time hashing.
The time bounds are optimal, matching input reading and explicit output writing.
Our floating-point prototype has lower sampling latency than the fastest applicable baseline in 17 of 18 PRODUCT settings on three queries over the STATS benchmark.
Construction savings on large joins can offset indexed materialization's faster sampling over a limited number of calls.
On a real-data path at $\mu=100$, our sampler achieves a $20.5\times$ speedup over indexed materialization for construction plus 300 calls, measured after common data preparation and loading.

\end{abstract}


\section{Introduction}
\label{sec:introduction}

Sampling a join should avoid paying for results that will be discarded.
Join-sampling algorithms use input statistics to generate uniform samples
directly~\cite{chaudhuri-motwani-narasayya-1999,zhao-christensen-li-hu-yi-2018},
while weighted sampling assigns different importance to
results~\cite{shekelyan-cormode-ma-shanghooshabad-triantafillou-2023}.
We study independent subset sampling, in which weights prescribe each
result's inclusion probability and all inclusion decisions are mutually independent.

Suppose an analyst wants to inspect
customer--order--item records. The analyst assigns importance weights in
$[0,1]$ to customers, orders, and items, and requests each joined record
with probability equal to the product of its three weights. Two records
may share a customer or an order, yet each receives an independent
inclusion decision, allowing one to be selected and the other omitted.
Sampling the customer table first would couple all records containing
that customer.

For a full join, which retains all query attributes, let $N$ count input
tuple occurrences and $m$ count relation occurrences. A witness
$t=(r_1,\ldots,r_m)$ combines one input tuple per relation occurrence to
produce a join result. Its PRODUCT inclusion probability is
$p(t)=\prod_{i=1}^{m}w(r_i)$, where $w(r_i)\in[0,1]$ is the weight
of input-tuple occurrence $r_i$. Write $\mathcal{J}$ for the witness set and
$J=|\mathcal{J}|$ for the full result size.

The sample size is random, with expectation
$\mu=\sum_{t\in\mathcal{J}}p(t)$. Inclusion indicators are mutually
independent Bernoulli variables with parameters $p(t)$, and repeated
calls return independent subsets. Section~\ref{sec:model} formalizes how
witness identities are preserved under set and bag semantics.

The challenge is to realize this distribution without processing all $J$
witnesses. Proportional sampling draws $t$ with probability $p(t)/\mu$,
but a fixed number of draws followed by duplicate removal generally changes
both the inclusion probabilities and their dependence. Alternatively,
materializing all witnesses enables an explicit subset index
with expected output-linear sampling time~\cite{bringmann-panagiotou-2017}.
Construction takes expected $O(N+mJ)$ time, and storing the witnesses
requires $O(mJ)$ words in addition to the input. A small expected sample
can thus require a large materialized index. This raises the question:
can we attain output-linear expected sampling time with
preprocessing and index space linear in $N$?

Esmailpour et al.~\cite{esmailpour-hu-huang-sintos-2026} support independent subset
sampling under PRODUCT and other weight aggregates. Their static PRODUCT
structure uses $O(N\log N\log\log N)$ preprocessing, $O(N\log N)$ space,
and expected $O(1+\mu\log N)$ time per sampling call for a fixed join
query. Bekkers et al.~\cite{bekkers-neven-pantelis-vansummeren-2026}
study Poisson sampling (independent Bernoulli inclusion) over acyclic
joins with probabilities supplied by one relation, whereas PRODUCT
combines weights across relations. For fixed full acyclic joins with static
PRODUCT probabilities, we remove the logarithmic factors from
preprocessing, index space, and sampling time simultaneously under the
real random-access machine (real RAM) model (Section~\ref{sec:model}).

\noindent\textbf{Our Approach.}
Our index groups input tuples by the attributes shared between adjacent
relations in a join tree~\cite{yannakakis-1981}. Within each group, the
total weight of subtree completions supports proportional draws through
alias sampling~\cite{walker-1977,vose-1991}, while the maximum completion
weight guides enumeration of high-probability witnesses.
Independent Bernoulli trials handle these witnesses directly.
A Poisson proposal process~\cite{gillespie-1976} draws from the entire
join, discards high-probability proposals, and thins the remaining proposals
by independently accepting each with an appropriate
probability~\cite{lewis-shedler-1979}. Deduplicating the accepted proposals
then produces the remaining independent inclusion decisions. Handling
high probabilities directly keeps the expected number of proposals
proportional to $\mu$.

The index uses expected $O(N)$ preprocessing and $O(N)$ words of index
space. Each call takes expected $O(1+m\mu)$ time, including writing the
$m$ tuple-occurrence identifiers of every returned witness.

\noindent\textbf{Guarantees and Empirical Findings.}
Our sampler is time-optimal on a real RAM with unit-cost exact
arithmetic and logarithms, continuous uniform randomness, and
expected-time hashing (Section~\ref{sec:model}). Preprocessing plus the
first call takes expected $O(N+1+m\mu)$ time, matching worst-case
input reading and expected output writing. Later calls match expected
output writing up to constant overhead. The $O(N)$ space bound covers the
index kept between calls. Output and working storage are separate.

On three queries over the STATS benchmark, our floating-point prototype
has lower sampling latency than the fastest applicable baseline with
complete measurements in 17 of 18 PRODUCT settings. In a separate paired
comparison on the real-data path NEWpath4 with real PRODUCT weights and
$\mu=100$, our sampler is $20.5\times$ faster than indexed materialization
for construction plus 300 calls, measured after common data preparation
and loading. Indexed materialization answers individual calls faster
in this comparison, so enough reuse can amortize its higher construction cost.
Small join results and dense samples can favor other methods.

Our contributions are:
\begin{itemize}
\item We design an index with expected linear preprocessing, linear index
space, and output-linear expected sampling time for independent PRODUCT
sampling over fixed full acyclic joins with static weights.
\item We prove the joint distribution of the returned subsets and time optimality for input reading and explicit output writing. Inclusion decisions are independent even for witnesses sharing input tuples, and repeated calls return independent subsets.
\item We demonstrate gains in total cost from avoiding materialization
and show how construction cost, sampling latency, and repeated calls
determine them.
\end{itemize}
Sections~\ref{sec:model}--\ref{sec:analysis} present the model, construction, and proofs. Section~\ref{sec:evaluation} evaluates the sampler.

\section{Problem and Computational Model}
\label{sec:model}

We consider a natural join query $Q$ that is fixed, full, and acyclic.
Full means that the output retains all query attributes. The $m$ relation
occurrences are positions in the query, so a self-join uses distinct
positions for repeated uses of the same relation. Acyclicity means that
$Q$ admits a join tree with one node per occurrence and the
running-intersection property: the nodes containing any attribute form a
connected subtree~\cite{yannakakis-1981}.

Let $D$ contain $N$ input tuple occurrences in total. A tuple occurrence
is a record instance, so equal-valued records can be distinct occurrences.
Each occurrence $r$ at node $v$ has a unique identifier (ID) and a weight
$w_v(r)\in[0,1]$. Weights stay fixed while the index is reused.
A witness $t=(r_1,\ldots,r_m)$ selects compatible input tuples, one per
node, and is represented by their vector of $m$ occurrence IDs.
Write $\mathcal{J}=Q(D)$ for the witness set and $J=|\mathcal{J}|$ for its
cardinality. Occurrence IDs distinguish equal-valued tuples under bag
semantics. Under set semantics, each full output determines its witness.

The PRODUCT inclusion probability of a witness and the expected sample size are
\begin{equation}
 p(t)=\prod_{v=1}^{m}w_v(r_v),\qquad
 \mu=\sum_{t\in\mathcal{J}}p(t).
 \label{eq:product}
\end{equation}
A sampling call returns $S\subseteq\mathcal{J}$ whose inclusion indicators
are mutually independent Bernoulli variables with parameters $p(t)$.
For $A\subseteq\mathcal{J}$, the probability of returning exactly $A$ is
\begin{equation}
 \Pr[S=A]=\prod_{t\in A}p(t)\prod_{t\in\mathcal{J}\setminus A}(1-p(t)).
 \label{eq:subset-law}
\end{equation}
The first product accounts for inclusion of all witnesses in $A$, and the
second for exclusion of all others. This is the PRODUCT subset law
(the joint distribution of $S$) studied by
Esmailpour et al.~\cite{esmailpour-hu-huang-sintos-2026}.
Each call uses fresh randomness, independent of earlier calls, so entire
returned subsets are independent across calls. In particular, inclusion
decisions remain independent for witnesses sharing an input tuple.
Outputs preserve the full occurrence vectors.

We measure preprocessing, index space, and sampling work separately.
Index space is the memory kept between sampling calls.
The cost of a sampling call includes writing all $m$ IDs of each returned witness.
We use a real random-access machine (real RAM) with unit-cost exact
arithmetic and comparisons, augmented with unit-cost logarithms,
continuous uniform random variables, and expected-time hash dictionaries
with exact collision resolution. We call this the exact-real model.
Unless a base is shown, $\log$ denotes the natural logarithm.

Query $Q$, attribute widths, and the sizes of keys on shared attributes
are fixed in our data-complexity analysis. Preprocessing constants may
depend on $Q$. We retain $m$ in sampling bounds to make the cost of
writing complete witnesses explicit.
Section~\ref{sec:evaluation} describes the floating-point prototype.

\noindent\textbf{Main Result.}
Under this model, our index uses expected $O(N)$ preprocessing and $O(N)$
words of index space. Each independent sampling call takes expected
$O(1+m\mu)$ time, including writing the full output vectors.
Theorem~\ref{thm:main} in Section~\ref{sec:analysis} gives the full
guarantee, with separate bounds for output and working storage.
Section~\ref{sec:index} builds the index for proportional draws and
high-probability witness enumeration. Section~\ref{sec:sampling} combines
these operations into independent subset sampling, handling high
probabilities with direct Bernoulli trials.

\section{A Linear Join Index}
\label{sec:index}

\begin{figure*}[t]
\centering
\includegraphics[width=0.98\textwidth]{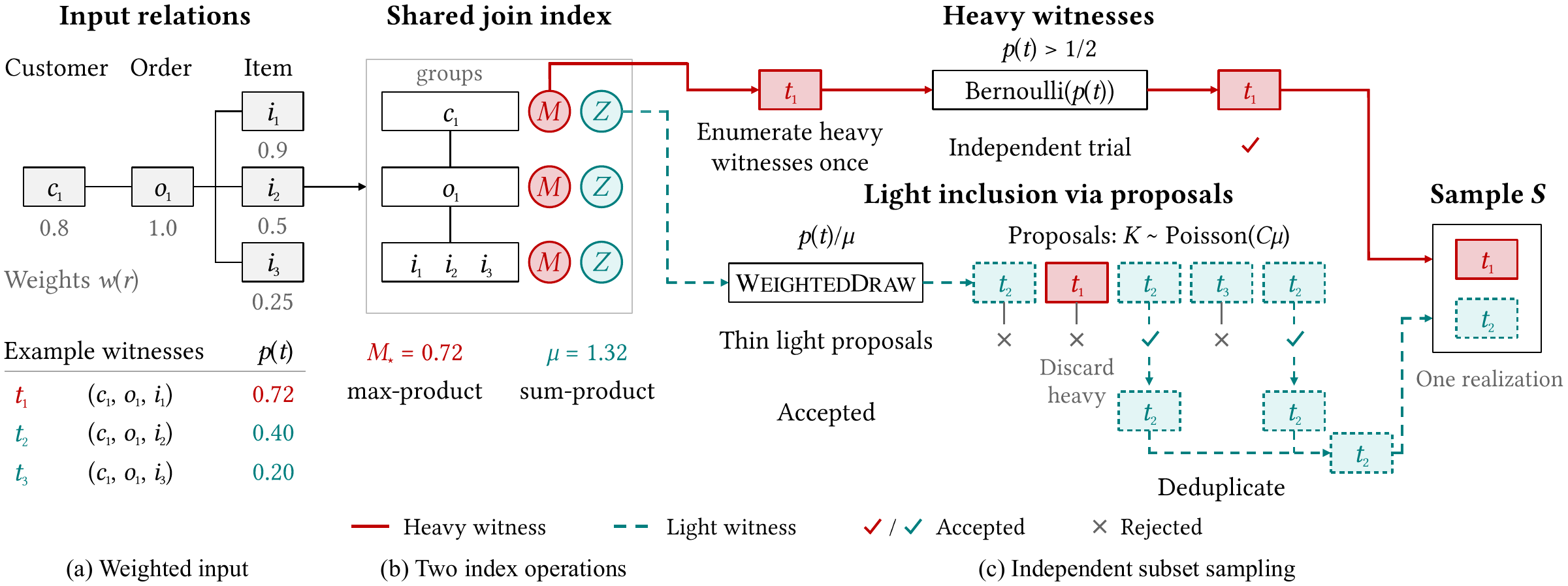}
\caption{One illustrative sampling call: (a) weighted input and example witnesses;
(b) the shared join index; (c) the two sampling paths and their combined output.
Red solid lines and teal dashed lines distinguish heavy and light witnesses.
Checkmarks and crosses mark
acceptance and rejection. The witnesses and proposal sequence illustrate the
process: proposals are handled one at a time, without materializing the join
or a candidate batch.}
\label{fig:overview}
\end{figure*}

Our index supports proportional draws over the full join and enumeration
of witnesses whose probabilities exceed a threshold.
Figure~\ref{fig:overview} follows the weighted input in (a), through the
shared index in (b), to one sampled subset in (c).
Total-weight summaries support the draws, while maximum-completion
summaries guide threshold enumeration. Section~\ref{sec:sampling}
combines these operations in the two sampling paths of Fig.~\ref{fig:overview}(c).

Root the join tree and fix a preorder $v_1,\ldots,v_m$ for the later
high-probability witness traversal. In the pseudocode, occurrence
vectors follow this preorder. For a nonroot node
$v$, group its tuples by their values on the attributes shared with the
parent. Let $G_v(s)$ be the group with separator key $s$.
The root has one group with an empty key. Empty separators also allow
disconnected components.

A subtree completion is a compatible choice of one tuple per node in a
subtree, weighted by the product of its tuple weights.
For completions with tuple $r$ fixed at $v$, $z_v(r)$ is the total weight
and $b_v(r)$ the maximum. The group summaries $Z_v(s)$ and $M_v(s)$
aggregate these tuple scores by sum and maximum over $G_v(s)$, respectively.

Let $s_{vc}(r)$ be the key that tuple $r$ supplies to child $c$.
Working bottom-up, we multiply $w_v(r)$ by the child summaries, then
aggregate over the tuple choices in each group:
\begin{equation}
\begin{aligned}
 z_v(r)&=w_v(r)\prod_{c\text{ child of }v}Z_c(s_{vc}(r)),\\
 Z_v(s)&=\sum_{r\in G_v(s)}z_v(r),
\end{aligned}\label{eq:sum-message}
\end{equation}
\begin{equation}
\begin{aligned}
 b_v(r)&=w_v(r)\prod_{c\text{ child of }v}M_c(s_{vc}(r)),\\
 M_v(s)&=\max_{r\in G_v(s)}b_v(r).
\end{aligned}\label{eq:max-message}
\end{equation}
Empty products equal one, and an empty group's sum and maximum are zero.
At the root, completions are full witnesses, so $Z_{v_1}(\varnothing)=\mu$.
We write $M_\star=M_{v_1}(\varnothing)$ for the largest witness probability.
In Fig.~\ref{fig:overview}, the root total is $0.72+0.40+0.20=1.32$,
and its maximum is $M_\star=0.72$.

In groups with positive total weight, build an alias table over the
positive $z_v$ scores. An alias table takes linear time to build and
supports exactly score-proportional draws in $O(1)$
time~\cite{walker-1977,vose-1991}. A binary max-heap over each group's
$b_v$ scores also takes linear time to build.
The sum-product construction follows classical weighted join
sampling~\cite{shekelyan-cormode-ma-shanghooshabad-triantafillou-2023}.
Max-product summaries also underlie ranked join
enumeration~\cite{tziavelis-ajwani-gatterbauer-riedewald-yang-2020}.
These heaps store tuple entries. The traversal in Section~\ref{sec:heavy}
uses them to select tuples and assemble high-probability witnesses.

For $\mu>0$, \textsc{WeightedDraw} selects a tuple from the root's alias
table, then recursively selects each child tuple from the group
determined by its parent. It returns the resulting full occurrence
vector $t$ and its product probability $p(t)$.
The construction uses linear index space and supports the required
proportional distribution.

\begin{lemma}
\label{lem:index}
The summaries, aliases, and heaps use $O(N)$ words of index space and expected
$O(N)$ construction time for fixed $Q$. If $\mu>0$, the procedure
\textnormal{\textsc{WeightedDraw}} returns $(t,p(t))$ with probability $p(t)/\mu$ in
$O(m)$ expected time.
\end{lemma}
\begin{proof}
Given a tuple, running intersection ensures that its children's conditioned
completions agree on every shared attribute. Their sums multiply, and
their maximizing completions combine into a consistent subtree completion.
Induction therefore proves both recurrences. Each tuple occurs in one parent group,
so group dictionaries, summaries, alias entries, and heap entries have
linear total size. Hash grouping and a bottom-up pass take expected linear
time for fixed $Q$.

The tuples selected by \textsc{WeightedDraw} are globally consistent by
running intersection. In the product of conditional selection
probabilities, a nonroot group's total $Z_v(s)$ appears in its parent's
tuple score and in the denominator of its own draw. These factors cancel,
leaving
\[
 \Pr[t]=\frac{\prod_v w_v(r_v)}{Z_{v_1}(\varnothing)}
        =\frac{p(t)}{\mu}.
\]
Only positive-score tuples are selected, so every visited child group has
positive total weight. There are $m$ alias selections and dictionary accesses.
Computing $p(t)$ and forming its ID vector also take $O(m)$ work.
\end{proof}

\section{Independent Subset Sampling}
\label{sec:sampling}

The conversion from proportional draws to independent inclusion uses
counts: a witness enters the subset when its accepted count is positive.
For a witness with inclusion probability $p{=}p(t){<}1$, a Poisson count with
mean $-\log(1-p)$ is positive with probability $p$~\cite{zhou-2015}.
A Poisson number of
proportional draws gives independent proposal counts for different
witnesses. Thinning, which independently accepts each proposal, can
convert these to the required accepted counts.

The needed count mean relative to $p$ grows without bound as $p$ approaches
one, making this conversion costly for high probabilities. We therefore
enumerate $H=\{t\in\mathcal{J}:p(t)>1/2\}$ separately and sample each
such \emph{heavy witness} with an independent Bernoulli trial
(Fig.~\ref{fig:overview}(c)). The remaining witnesses are light and use
the count construction. Each heavy witness contributes more than $1/2$
to $\mu$, so $|H|\leq2\mu$ (strictly when $\mu>0$).
Enumerating $H$ with work proportional to its size thus keeps this path
within the target sampling bound.

\subsection{Enumerating Heavy Witnesses}
\label{sec:heavy}

A preorder prefix fixes the first few tuple choices in the order
$v_1,\ldots,v_m$. Let $\beta$ be the largest weight of any full witness
consistent with those choices. The unassigned part consists of
\emph{frontier subtrees} attached to the selected prefix, and multiplying
their conditional maxima by the selected tuple weights gives $\beta$.
When $\beta>0$, running intersection lets us combine the
maximizing completions into a legal full witness of weight $\beta$. In particular, a prefix with $\beta>1/2$ has a heavy completion.

At the next node $v$, the assigned parent determines a group $G_v(s)$
whose maximum $M=M_v(s)$ is one factor in $\beta$.
Choosing tuple $r$ replaces $M$ by $b_v(r)$ while leaving all other
factors unchanged, giving
\begin{equation}
 \beta'=\beta\,b_v(r)/M,\qquad
 \beta'>1/2\ \Longleftrightarrow\ b_v(r)>M/(2\beta).
 \label{eq:prefix-update}
\end{equation}
Only prefixes with $\beta>1/2$ are expanded, so $M>0$ and division is safe.
The helper $\textsc{HeapAbove}(G,\theta)$ traverses the binary max-heap
inside group $G$ in depth-first order. It yields a tuple entry if its
$b_v$ score exceeds $\theta$, then visits its heap children.
Otherwise it prunes that heap subtree. The outer preorder traversal
combines these entries into complete witnesses.
Each recursive call retains $\beta$ in its caller's frame, so a tuple
choice needs only the update in~\eqref{eq:prefix-update}.
We next bound the work of reporting every heavy witness exactly once.

\begin{lemma}
\label{lem:heavy}
The preorder traversal using \eqref{eq:prefix-update} enumerates exactly
$H$, once each, in expected $O(1+m|H|)$ time.
\end{lemma}
\begin{proof}
Initially $\beta=M_\star$. Selecting a tuple replaces exactly one frontier
factor, proving the maximum-completion invariant inductively. All other
frontier factors remain in $\beta$, including those needed after returning
from a sibling subtree. Every passing choice has a heavy completion.
Conversely, each prefix of a heavy witness passes the strict threshold.
At depth $m$, $\beta=p(t)$, so the traversal reaches precisely the heavy
witnesses. Each occurrence vector follows a unique sequence of tuple choices.

At any fixed prefix depth, distinct passing prefixes have disjoint sets
of completions, each containing a heavy witness. There are therefore at
most $|H|$ passing prefixes per depth and at most $m|H|$ nonempty passing
prefixes altogether. Within a group, a heap walk with $\ell$ passing nodes
examines at most $1+2\ell$ nodes: every examined nonroot heap node is a child
of a passing node. A nonterminal passing prefix has $\ell\geq1$, since the
heap root has score $M$ and passes the threshold. Failed heap tests
therefore add only a constant factor.
Maintaining the prefix and saved bound takes constant work per choice,
and each group lookup takes expected constant time. The deterministic bound on visited prefixes and heap nodes, together with
these lookups, proves the expected time bound. Writing complete vectors
costs $O(m|H|)$. When $M_\star\leq1/2$, one
comparison establishes that $H$ is empty.
\end{proof}

\subsection{Sampling Light Witnesses}

For a light witness with probability $p$, let $f(p)$ be the multiplier
that converts $p$ to the target Poisson mean $-\log(1-p)$.
We use a common upper bound $C$ on these multipliers to set the proposal rate:
\begin{equation}
\begin{aligned}
 f(x)&=\frac{-\log(1-x)}{x}\ (0<x<1),\quad f(0)=1,\\
 C&=f(\min\{M_\star,1/2\}).
\end{aligned}
 \label{eq:envelope}
\end{equation}

The identity $f(x)=\int_0^1(1-xs)^{-1}\,ds$ shows that $f$ is increasing.
Consequently, $f(p(t))/C\leq1$ for every light witness, giving a valid
acceptance probability. Also, $1\leq C\leq2\log2$, so a proposal count
$K\sim\operatorname{Poisson}(C\mu)$ has expectation at most
$2\log2\,\mu$. Thus the proposals incur only constant-factor overhead
relative to $\mu$.

We generate unit-rate Poisson arrivals over the simulated interval
$[0,C\mu]$ using independent exponential gaps $-\log U$, where $U$ is
uniform on $(0,1)$. Simulated time is an algorithmic device and requires
no real-time waiting.
At each arrival, call \textsc{WeightedDraw} over the full join
(Fig.~\ref{fig:overview}(c)). Reject heavy proposals and
accept each light proposal independently with probability $f(p(t))/C$.
Deduplicate only the accepted light witnesses: in Fig.~\ref{fig:overview}(c),
the two accepted $t_2$ proposals yield one returned occurrence vector.
We use Poisson simulation
and thinning~\cite{gillespie-1976,lewis-shedler-1979} with the weighted
join index supplying the arrival marks.

Algorithm~\ref{alg:subset} uses a temporary ID vector $a$, allocated only
when needed. \textsc{WeightedDraw} uses the proportional sampler of
Section~\ref{sec:index} to fill this buffer in preorder and returns it
with its product probability.
\textsc{HeapAbove} performs the thresholded heap traversal described above,
and \textsc{AppendCopy} appends an owned copy of the occurrence vector
to the output. Within a sampling call, \textsc{VisitHeavy} shares
$S$ and $a$ with \textsc{SubsetSample}.
The dictionary $D_L$ stores accepted light ID vectors, resolving hash
collisions by comparing full vectors. Insertion returns true precisely
for a new key, storing a copy or a reference to its output vector.

The variable $h$ records cumulative simulated arrival time.
Recursive calls pass $\beta$ by value, retaining the previous bound for
the next tuple choice. Every uniform draw is fresh and independent, as
noted in the pseudocode.

\begin{algorithm}[t]
\caption{Independent PRODUCT Subset Sampling}
\label{alg:subset}
\algrenewcommand{\algorithmicindent}{1em}
\compactalgorithmic
\begin{algorithmic}[1]
\Require Grouped index, $\mu,M_\star,C$, preorder $v_1,\ldots,v_m$
\Ensure Independent subset of occurrence vectors
\Statex{\small\color[gray]{.4}\text{//} Each $U$ denotes a fresh independent $\mathrm{Uniform}(0,1)$ draw.}
\Function{SubsetSample}{}
  \State $S\gets[\,]$, $D_L\gets\varnothing$, $a\gets\mathrm{unallocated}$
  \If{$\mu=0$} \State \Return $S$ \EndIf
  \If{$M_\star>1/2$}
    \State Allocate $a[1..m]$ \Comment{Scratch only when needed}
    \State \Call{VisitHeavy}{$1,M_\star$} \Comment{Heavy witnesses}
  \EndIf
  \State $h\gets-\log U$ \Comment{First simulated arrival}
  \While{$h\leq C\mu$}
    \If{$a$ is unallocated} \State Allocate $a[1..m]$ \EndIf
    \State $(t,p)\gets\Call{WeightedDraw}{a}$
    \If{$p\leq1/2$}
      \If{$U<f(p)/C$}
        \If{$D_L.\mathrm{insert}(t)$} \Comment{Deduplicate accepted IDs}
          \State \Call{AppendCopy}{$S,t$}
        \EndIf
      \EndIf
    \EndIf
    \State $h\gets h-\log U$
  \EndWhile
  \State \Return $S$
\EndFunction
\vspace{1mm}
\Procedure{VisitHeavy}{$i,\beta$}
  \If{$i=m+1$}
    \If{$U<\beta$} \State \Call{AppendCopy}{$S,a$} \EndIf
    \State \Return
  \EndIf
  \State $v\gets v_i$, $s\gets$ parent key from $a$ (empty at root)
  \State $M\gets M_v(s)$
  \For{$r\in\Call{HeapAbove}{G_v(s),M/(2\beta)}$}
    \State $a[i]\gets\mathrm{ID}(r)$
    \State \Call{VisitHeavy}{$i+1,\beta b_v(r)/M$} \Comment{Update maximum}
  \EndFor
\EndProcedure
\end{algorithmic}
\end{algorithm}

\section{Correctness and Complexity}
\label{sec:analysis}

We first establish the joint distribution of the returned subset,
including independence within and across calls, and then combine this
result with the index and enumeration bounds to prove
Theorem~\ref{thm:main}.

\begin{lemma}
\label{lem:law}
Algorithm~\ref{alg:subset} satisfies \eqref{eq:subset-law}, including mutual
independence of all witness indicators and independence across calls.
\end{lemma}
\begin{proof}
First suppose $\mu>0$. Let $K$ be the number of arrivals. Exponential gaps
give $K\sim\mathrm{Poisson}(C\mu)$. By Lemma~\ref{lem:index}, each arrival
is independently marked $t$ with probability $p(t)/\mu$. Poisson splitting
therefore gives mutually independent Poisson proposal counts with means
$Cp(t)$. To verify this joint independence, let $x_t$ be the formal
variable for witness $t$'s proposal count. The joint
probability-generating function is
\[
\begin{aligned}
 &\exp\!\left(C\mu\left[\sum_t\frac{p(t)}{\mu}x_t-1\right]\right)\\
 &\hspace{2em}=\prod_t\exp\!\left(Cp(t)(x_t-1)\right),
\end{aligned}
\]
which factors into the individual Poisson generating functions.

Independent thinning of a light witness's proposals gives a Poisson
accepted count $A_t$ with mean
\[
 Cp(t)\frac{f(p(t))}{C}=-\log(1-p(t)).
\]
The accepted counts remain mutually independent. Deduplication applies the map
$A_t\mapsto\mathbf{1}[A_t>0]$ separately to each count, yielding independent
indicators with
$\Pr[A_t>0]=1-\exp(\log(1-p(t)))=p(t)$.
Note that every proposal receives its thinning decision before accepted occurrences
are deduplicated, as required by this count calculation.

Lemma~\ref{lem:heavy} visits every heavy witness exactly once. Its independent
Bernoulli coin has parameter $p(t)=\beta$ and is independent of the entire
arrival process, proving the joint law for heavy and light witnesses
together.
A zero-probability witness is never proposed and is absent.
A unit-probability witness is heavy and included deterministically, so no
logarithm is evaluated at $1-p=0$. The boundary $p=1/2$ is light. If
$\mu=0$, including when the join is empty, every required indicator is zero and the immediate return is correct.

Finally, each call uses a fresh
dictionary and independent arrival, alias, thinning, and heavy-coin
randomness, which makes the returned subsets independent across calls.
\end{proof}

The preceding lemmas establish the index, enumeration, and distribution
guarantees. We now combine them to bound the full sampling call's time
and space.

\begin{theorem}
\label{thm:main}
For every fixed full acyclic join $Q$, an index with expected $O(N)$
preprocessing time and $O(N)$ words of index space supports the law in
\eqref{eq:subset-law} in expected $O(1+m\mu)$ time per independent sampling call.
The returned vectors and deduplication state occupy $O(m|S|)$ words,
separately from the index. Each sampling call uses an additional
$O(m\log(N+1))$ words of working space, excluding the returned vectors
and deduplication state.
\end{theorem}
\begin{proof}
Lemma~\ref{lem:index} bounds preprocessing and index space.
Lemma~\ref{lem:law} proves correctness.
By Lemma~\ref{lem:heavy}, heavy enumeration and coins cost expected
$O(1+m|H|){=}O(1+m\mu)$. We use $K$ weighted draws and $K{+}1$
exponential gaps, including the final gap beyond $C\mu$. Since
$\mathbb{E}[K]=C\mu\leq2\log2\,\mu$, proposal generation, probability
calculation, hashing, and collision-resolving comparisons cost
expected $O(1+m\mu)$. Copying returned vectors costs expected
$O(m\mathbb{E}|S|)=O(m\mu)$.

When $m$ is explicit and $\mu$ is small, unconditional initialization
of an $m$-entry vector can exceed the stated $O(1+m\mu)$ bound.
Allocating scratch only when needed avoids this cost.
A heavy traversal begins only if $M_\star>1/2$, implying
$\mu>1/2$, so its $O(m)$ setup is chargeable to $m\mu$. Otherwise scratch
is allocated only on the first arrival, whose probability is
$1-e^{-C\mu}\leq C\mu$. Its expected initialization cost is again
$O(m\mu)$. Heap stacks allocate entries as traversal visits them. Empty dictionaries
require constant setup, and releasing each call's dictionary takes work
proportional to its accepted entries.

The accepted light IDs and output occupy $O(m|S|)$ words. The current vector and join recursion take $O(m)$ scratch
words. Each suspended binary-heap traversal has depth $O(\log(N+1))$, and
at most $m$ traversals are suspended, giving the claimed additional space.
Each arrival is processed immediately and discarded.
\end{proof}

\noindent\textbf{Time Optimality.}
The time bounds match the required input and output work in this model.
Writing explicit occurrence vectors requires expected $\Omega(m\mu)$
operations. For input reading, even a one-relation all-zero instance forces
every potential weight to be inspected before its first exact answer:
changing an uninspected weight to one would require a different deterministic
answer. Preprocessing plus the first sampling call therefore requires
$\Omega(N)$ input reading in the worst case. Input reading and output writing
thus account for the $O(N+1+m\mu)$ cost through the first call, and explicit
output writing accounts for the $O(1+m\mu)$ cost of later calls up to constant
overhead. These time bounds are optimal under our interface. The space guarantee
is the $O(N)$ upper bound proved above, with output and scratch accounted
for separately.

\section{Experimental Evaluation}
\label{sec:evaluation}

We evaluate how the theoretical cost bounds translate into sampling time,
construction time, and memory use. Real-data and controlled workloads
test how weight distributions, input and join sizes, and sample density
affect the benefit, while a paired materialization study asks how many
calls 
amortize a larger index construction cost. We also
check the floating-point prototype's output against the prescribed subset
distribution.

\subsection{Experimental Setup}

\noindent\textbf{Methods.}
Our sampler implements Sections~\ref{sec:index}--\ref{sec:sampling} in C++20
with double-precision floating-point (binary64)
probabilities~\cite{ieee-754-2019} and a
pseudorandom number generator (PRNG).
A sampling call returns one subset for a fixed join query $Q$.
Table~\ref{tab:methods} summarizes our baseline implementations, which we
have open-sourced together with the sampler for
reproducibility~\cite{join-subset-sampling}.
The accompanying artifact provides reproduction instructions and records supporting
all reported experimental results, including those discussed only
in the text.

We compare against \textsc{Score}, our implementation of the PRODUCT
subset sampler of Esmailpour et al.~\cite{esmailpour-hu-huang-sintos-2026}.
It groups witnesses
by total score and uses a probability upper bound for each bucket. With quantization step
$\delta=1/m$, a tuple of positive weight $w$ contributes
$\lfloor-\log_2(w)/\delta\rfloor$, and summaries count subtree completions
by total score. Exact convolution combines these count arrays across
child subtrees.
A rank is a witness's position in a fixed enumeration of the join or a
score bucket. Geometric skipping jumps over Bernoulli failures to the next
candidate, with success probability equal to the relevant probability upper bound.
Alias tables select score buckets, and geometric gaps select candidate
ranks within them. Thinning reduces a bucket's upper bound to each
witness's probability.

\textsc{Global} and \textsc{Root} test how the tightness of probability
upper bounds affects candidate generation and rejection. We call these
bounds \emph{envelopes} and their sum over witnesses \emph{envelope mass}.
\textsc{Global} draws a binomial candidate count. A root block contains all
witnesses sharing a root tuple $r$. Let $J_r$ count these witnesses and
$U_r$ bound their probabilities. \textsc{Root} independently selects each
block with probability $1$$-$$(1$$-$$U_r)^{J_r}$, the probability of at least one
candidate, then draws its count from a $\operatorname{Binomial}(J_r,U_r)$
distribution conditioned on being positive. Both generate a
uniform subset of distinct candidates, using repeated uniform draws in
sparse cases or fixed-size selection during enumeration in dense cases,
and thin by the ratio of true probability to its envelope.
\textsc{Root} is especially effective when one relation carries the
probability attribute, the scope studied by Bekkers
et al.~\cite{bekkers-neven-pantelis-vansummeren-2026}. We implement the
bound-based sampling mechanism as a baseline.

\textsc{Rank} is a specialized comparison for uniform probabilities,
using geometric skipping over a fixed join enumeration.
\textsc{Scan} stores the join and performs an independent Bernoulli test
on every witness in each call.
\textsc{Materialized} tests whether faster repeated sampling can repay
the cost of materialization. It stores all $mJ$ occurrence IDs in a flat
array and builds an independent-subset index following
Bringmann--Panagiotou~\cite{bringmann-panagiotou-2017}. Probability-one
witnesses are stored separately, and positive probabilities below one are
grouped under dyadic upper bounds, with a common bound for the tail.
Alias tables select buckets contributing candidates, geometric skipping
selects candidates within a bucket, and thinning corrects their
probabilities. Calls copy selected full ID vectors.

\begin{table}[t]
\centering
\caption{Implemented methods. Every method returns complete occurrence vectors.}
\label{tab:methods}
\small
\begin{tabular}{@{}>{\columncolor{white}[0pt][\tabcolsep]}l>{\columncolor{white}[\tabcolsep][0pt]}p{0.71\columnwidth}@{}}
\toprule
Method & Index and Sampling Mechanism \\
\midrule
\rowcolor{rowshade} \textsc{Ours} & Weighted aliases, maximum-completion heaps, Poisson proposals.\\
\textsc{Score} & Total-score buckets, exact convolution, rank access and thinning.\\
\rowcolor{rowshade} \textsc{Global} & Envelope mass $J\max_t p(t)$, binomial count and uniform candidates.\\
\textsc{Root} & Sampling within root blocks, envelope mass $\sum_r J_r U_r$.\\
\rowcolor{rowshade} \textsc{Rank} & Geometric skipping over a fixed enumeration, for uniform probabilities.\\
\textsc{Scan} & Stored witnesses, independent Bernoulli test of every witness.\\
\rowcolor{rowshade} \textsc{Materialized} & Stored witnesses with an explicit subset index.\\
\bottomrule
\end{tabular}
\unskip
\end{table}

\noindent\textbf{Workloads.}
We use the simplified STATS-CEB snapshot~\cite{han-et-al-2022-stats-ceb}
and May-2013 IMDb data from the Join Order Benchmark
(JOB)~\cite{leis-et-al-2015-job}. 
Benchmark queries provide
real query structures. Two additional STATS queries test large implicit
results from joins on nonunique keys, while JOB queries contrast
larger inputs with smaller, explicitly storable results.
Table~\ref{tab:workloads} lists input size $N$ after local filtering and
full join cardinality $J$.

Q010 joins
Comments--Users--Posts. Q064 links Posts to PostLinks, PostHistory, and
Comments, then joins Comments--Users--Votes. Both use the original benchmark
predicates. NEWpath4 is an unfiltered PostHistory--Votes--Comments--PostLinks
path, joined successively on PostId, UserId, and PostId. We preselected it
to test a path between fact tables whose leaf joins use nonunique keys.
NEWbranch5 adds a Votes--Badges edge on UserId to NEWpath4,
forming a branch of five unfiltered fact relations with about 5.4 billion
witnesses.

JOB1a uses its five relations and original local predicates.
JOB17a uses seven relations, including cast\_info,
movie\_companies, and movie\_keyword, with its original predicates on names
beginning with B, US companies, and the keyword
\texttt{character-name-in-title}. For both JOB queries, we replace the
final aggregate with complete vectors of contributing tuple IDs.
NEWbranch5 and JOB17a were selected by structure before inspecting
cardinalities and measured in an independent extension batch.

Data preparation preserves SQL NULL semantics and removes nonmatching NULL
keys. It applies local filters only, with every relation node preserved and
join-driven semijoin pruning and primary-key folding disabled.

\begin{table}[t]
\centering
\caption{Real join workloads. Original STATS queries have 12 probability cases each, and JOB1a has 10 after identical scales are merged. Each extension query has five cases.}
\label{tab:workloads}
\small
\begin{tabular}{@{}>{\columncolor{white}[0pt][\tabcolsep]}lrr>{\columncolor{white}[\tabcolsep][0pt]}r@{}}
\toprule
Query & $m$ & $N$ & $J$\\
\midrule
\rowcolor{rowshade} Q010 & 3 & 221,401 & 780,683\\
Q064 & 6 & 546,248 & 692,609\\
\rowcolor{rowshade} NEWpath4 & 4 & 520,548 & 20,181,980\\
JOB1a & 5 & 3,937,238 & 142\\
\midrule
\rowcolor{rowshade} NEWbranch5 & 5 & 600,399 & 5,415,045,576\\
JOB17a & 7 & 46,333,958 & 258,289\\
\bottomrule
\end{tabular}
\unskip
\end{table}

Controlled workloads isolate four effects. The size family increases
input size using ten six-relation chains and branches, varying the
number of separator keys $K=128,512,2048,8192,32768$
at degree $d=5$, where every separator-key value occurs in $d$ rows on
each side of every join-tree edge. Their sizes are $N=6Kd$ and $J=Kd^6$.
The degree family increases join size at fixed input size: five four-relation chains
hold $N=65{,}536$ fixed and increase $d=2,4,8,16,32$, giving
$J=16{,}384d^3$. Eleven triangular cases isolate loose probability bounds using
$A(x),B(x,y),C(y,z),D(z)$ with $0\le x\le y\le z<K$ and
$K=16,32,\ldots,1024$. Here \emph{triangular} describes the ordered-key
constraint in the data, while the join remains an acyclic path.
Finally, five density cases on this path at $K=64$ increase the fraction
of results to be returned.

\noindent\textbf{Probability Settings.}
We control expected output size by scaling weights downward.
For unscaled expected sample size $\mu_0$, we multiply every tuple weight
in one selected relation by $\alpha=\min\{1,\tau/\mu_0\}$, with targets
$\tau=1,100,10{,}000$. Comparisons use the actual $\mu$ when the unscaled
instance cannot reach a target.

The original real-data queries use four modes for different comparisons.
Uniform probabilities admit the specialized \textsc{Rank} sampler.
One-relation weights test probabilities concentrated in a single relation.
Real PRODUCT weights combine features across relations, while shuffled
PRODUCT weights test their association with the join keys.
Independent shuffling preserves each filtered relation's weight
multiset but changes how weights are associated with join keys and matching
tuples.

Features represent recency and numerical importance. With $\epsilon=1/64$,
date weights linearly map a nonconstant CreationDate domain's
minimum and maximum in the original table to $[\epsilon,1]$.
Let $q_{99}$ be the original table's nearest-rank 99th percentile of a
nonnegative feature $x$. For $q_{99}>0$, numerical importance is
$\epsilon+(1-\epsilon)\min\{1,\log(1+x)/\log(1+q_{99})\}$ for nonnegative
$x$.
Q010 combines comment recency, post ViewCount, and user
Reputation. Q064 combines four CreationDate features, post FavoriteCount,
and user Reputation. NEWpath4 uses recency in all four relations. The
one-relation modes weight Votes for NEWpath4 and Users for Q010 and Q064.
JOB1a uses title production-year recency over 1880--2019 and
$\epsilon+(1-\epsilon)/\rho$ for valid top-250 ranks $\rho$, assigning $\epsilon$
to other rank subtypes.
All original STATS cases reach the three targets.
JOB1a's maximum expected sample sizes are approximately $142$, $113.6$,
$5.02$, and $1.79$ in the four modes, so some settings stop below their
target. Identical unscaled PRODUCT settings are included once.

NEWbranch5 uses the path's four date features and Badges.Date with the same
mapping. JOB17a combines title-year recency with
$\epsilon+(1-\epsilon)/\rho$ for positive cast billing order $\rho$ (otherwise
$\epsilon$). Its other five relations have weight one. Each extension query
has PRODUCT targets $1,100,10{,}000$ and uniform and one-relation settings at
$100$, all reached. The single weighted relations are Votes and title,
respectively.

The controlled size and degree families use independent weights $2^{-u}$,
with $u$ uniform on $(0,3)$, normalized downward to $\mu=100$. In the
triangular family, only $A(0)$ and $D(K-1)$ have endpoint weight one.
Other endpoint weights are $K^{-2}$, and middle weights are one. We include
unscaled instances and distinct target-100 instances. The unscaled global
and root envelope masses are $\Theta(K^3)$ and $\Theta(K^2)$, respectively,
with $\mu=\Theta(K)$. Its primary-key leaves make this a synthetic case
for isolating loose probability bounds. Dense cases have constant output
probabilities $1/1024,1/16,1/2,3/4,1$.

\begin{figure*}[t]
	\centering
	\includegraphics[width=0.94\textwidth]{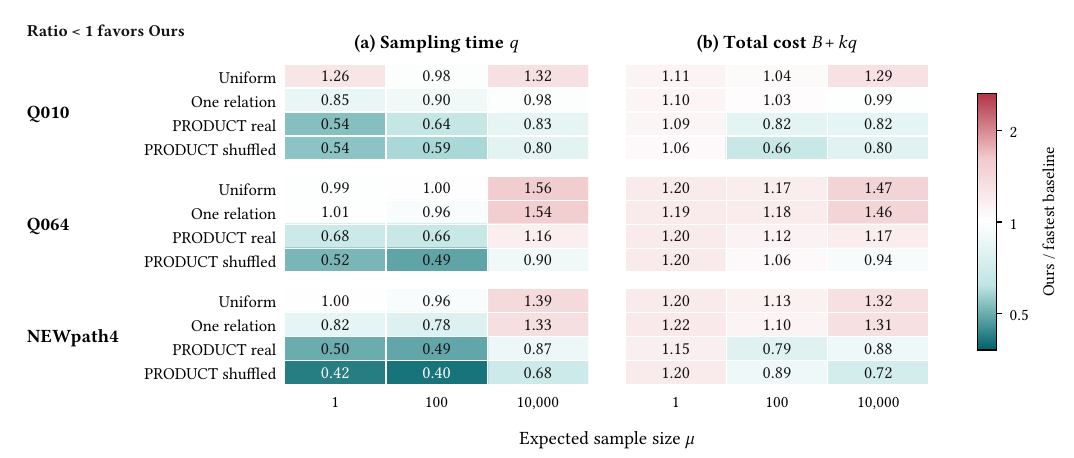}
	\vspace{-1mm}
	\caption{All 36 original STATS cases. Cells show \textsc{Ours} divided by the fastest applicable baseline with complete measurements for each metric. Teal indicates ratios below one and red above one on a logarithmic color scale. Columns $\mu=1,100,10{,}000$ use $k=1000,300,100$, respectively. Values are ratios of four-process medians rounded to two decimal places.}
	\label{fig:stats}
\end{figure*}

\noindent\textbf{Measurement Protocol.}
Let $B$ be index construction time and $q$ the average time per measured
sampling call within a process. For $k$ measured calls, the \emph{total
method cost} $B+kq$ includes construction and sampling after common data
preparation and loading. Call timing includes writing all occurrence IDs
and constructing and destroying the returned subset.
Real-data targets $1,100,10{,}000$ use
$k=1000,300,100$, with $20,5,2$ warmup calls. Controlled target-100,
unscaled triangular, and dense cases use $300,100,20$ measured calls and
$5,2,2$ warmups. Every configuration in Tables~\ref{tab:method-results}
and~\ref{tab:extension-results} uses 300 measured calls per process.

Each method and case has four process repeats, with rotated primary-method
order and distinct seeds. We report medians and observed ranges.
The paired \textsc{Materialized} study alternates the two methods and
computes ratios from corresponding repeats. Those comparisons
use medians of paired ratios, while the other comparisons use ratios of
four-process medians.

All methods receive the same prepared tuples and rooted join tree.
External data preparation
filters and encodes data, computes $\mu_0$ and normalization, and selects
a common root with approximately minimum envelope mass.
Each process then loads prepared rows, keys, and weights and
roots the tree. Construction $B$ starts from this common input. Each method
builds its own counts, sums, maxima, aliases, or score summaries.
For \textsc{Materialized}, $B$ includes completion counts, full enumeration,
allocation of the flat ID array, and construction of the subset index.
Temporary counts and the probability array are released during construction.
Preparation and loading are recorded outside
$B+kq$, and changed data or weights can require repeating preparation.
Adding these common stages to both methods would move cost ratios toward
one. Allocated index memory estimates the capacity kept between calls.
Peak resident set size (RSS) measures memory across the whole process.
We report memory in MiB, using GiB for process limits.

We compile all methods with GCC 11.4 using \texttt{-O3 -DNDEBUG} on Ubuntu 22.04.
Runs execute serially, pinned to one core of a Xeon Gold 6342 server.
The host is not isolated. The default limits are 8\,GiB of address space
and 120\,s per process. For JOB17a's larger input, we fixed 32\,GiB and
600\,s limits before timing. All studies use the same CPU, four repeats,
and definitions of $B$, $q$, and call counts.

The original comparisons use \textsc{Score}, \textsc{Global}, and
\textsc{Root}, with \textsc{Rank} for uniform probabilities and
\textsc{Scan} for $J\le10^6$. The separate paired study covers all 36 STATS
cases, ten JOB1a cases, and 21 controlled cases within its 25-million-output
budget. In the extension, \textsc{Scan} and \textsc{Materialized} run on
JOB17a but are omitted from NEWbranch5 because it exceeds their output
budgets. \textsc{Rank} again applies only to uniform probabilities.
For each timing metric, the comparison baseline is the fastest applicable
baseline with complete measurements, so its identity can differ across
configurations and between sampling time and total cost. Timeouts and
memory exhaustion leave $B$ and $q$ unmeasured and exclude the affected
method from these ratios. We report the affected configurations with
their results below.

\subsection{Sampling Performance}

\begin{table*}[t]
\centering
\caption{Construction time $B$ (ms) and sampling time $q$ ($\mu$s) in three original batches and an independently measured paired batch. Entries show four-process medians with full observed ranges $R=\max-\min$ in parentheses. Bold marks the lowest median per metric and batch. STATS, JOB1a, and paired cases use real PRODUCT weights. A dash marks a method omitted from that batch. NA: \textsc{Scan} exceeds its $J\le10^6$ output budget. TO: all four processes reached 120\,s, leaving $B$ and $q$ unmeasured. \textsc{Rank} applies only to uniform probabilities.}
\label{tab:method-results}
\fontsize{9}{10.8}\selectfont
\setlength{\tabcolsep}{4.5pt}
\begin{tabular}{@{}lcccccccc@{}}
\toprule
\textbf{Method} & \multicolumn{2}{c}{\shortstack{\textbf{Original STATS}\\NEWpath4\\$\mu=100$}} & \multicolumn{2}{c}{\shortstack{\textbf{Original controlled}\\Triangular $K=256$\\$\mu=100$}} & \multicolumn{2}{c}{\shortstack{\textbf{Original JOB1a}\\PRODUCT real\\Actual $\mu=5.017566$}} & \multicolumn{2}{c}{\shortstack{\textbf{Paired batch}\\NEWpath4\\$\mu=100$}}\\
\cmidrule(lr){2-3}\cmidrule(lr){4-5}\cmidrule(lr){6-7}\cmidrule(lr){8-9}
 & $B$ (ms) & $q$ ($\mu$s) & $B$ (ms) & $q$ ($\mu$s) & $B$ (ms) & $q$ ($\mu$s) & $B$ (ms) & $q$ ($\mu$s)\\\midrule
\textsc{Ours} & \makebox[24.75pt][r]{70.97}\hspace{1.4pt}\makebox[21.875pt][l]{\rangeval{1.86}} & \makebox[24.75pt][r]{\textbf{85.26}}\hspace{1.4pt}\makebox[18.125pt][l]{\rangeval{1.11}} & \makebox[20.25pt][r]{6.39}\hspace{1.4pt}\makebox[18.125pt][l]{\rangeval{0.32}} & \makebox[29.25pt][r]{37.57}\hspace{1.4pt}\makebox[21.875pt][l]{\rangeval{1.64}} & \makebox[29.25pt][r]{1379.02}\hspace{1.4pt}\makebox[21.875pt][l]{\rangeval{13.72}} & \makebox[20.25pt][r]{3.16}\hspace{1.4pt}\makebox[18.125pt][l]{\rangeval{0.19}} & \makebox[29.25pt][r]{\textbf{71.34}}\hspace{1.4pt}\makebox[25.625pt][l]{\rangeval{1.19}} & \makebox[20.25pt][r]{84.87}\hspace{1.4pt}\makebox[18.125pt][l]{\rangeval{4.42}}\\[1.4pt]
\textsc{Score} & \makebox[24.75pt][r]{476.90}\hspace{1.4pt}\makebox[21.875pt][l]{\rangeval{10.04}} & \makebox[24.75pt][r]{208.17}\hspace{1.4pt}\makebox[18.125pt][l]{\rangeval{1.42}} & \makebox[20.25pt][r]{34.88}\hspace{1.4pt}\makebox[18.125pt][l]{\rangeval{1.37}} & \makebox[29.25pt][r]{\textbf{30.35}}\hspace{1.4pt}\makebox[21.875pt][l]{\rangeval{1.54}} & \makebox[29.25pt][r]{5088.43}\hspace{1.4pt}\makebox[21.875pt][l]{\rangeval{54.80}} & \makebox[20.25pt][r]{3.63}\hspace{1.4pt}\makebox[18.125pt][l]{\rangeval{0.19}} & -- & --\\[1.4pt]
\textsc{Global} & \makebox[24.75pt][r]{\textbf{58.45}}\hspace{1.4pt}\makebox[21.875pt][l]{\rangeval{0.65}} & \makebox[24.75pt][r]{278.99}\hspace{1.4pt}\makebox[18.125pt][l]{\rangeval{5.31}} & TO & TO & \makebox[29.25pt][r]{\textbf{1216.43}}\hspace{1.4pt}\makebox[21.875pt][l]{\rangeval{12.43}} & \makebox[20.25pt][r]{49.62}\hspace{1.4pt}\makebox[18.125pt][l]{\rangeval{0.64}} & -- & --\\[1.4pt]
\textsc{Root} & \makebox[24.75pt][r]{70.60}\hspace{1.4pt}\makebox[21.875pt][l]{\rangeval{0.79}} & \makebox[24.75pt][r]{174.14}\hspace{1.4pt}\makebox[18.125pt][l]{\rangeval{7.10}} & \makebox[20.25pt][r]{\textbf{4.36}}\hspace{1.4pt}\makebox[18.125pt][l]{\rangeval{0.12}} & \makebox[29.25pt][r]{5248.67}\hspace{1.4pt}\makebox[21.875pt][l]{\rangeval{71.55}} & \makebox[29.25pt][r]{1218.06}\hspace{1.4pt}\makebox[21.875pt][l]{\rangeval{34.25}} & \makebox[20.25pt][r]{2.86}\hspace{1.4pt}\makebox[18.125pt][l]{\rangeval{0.13}} & -- & --\\[1.4pt]
\textsc{Scan} & NA & NA & NA & NA & \makebox[29.25pt][r]{1707.86}\hspace{1.4pt}\makebox[21.875pt][l]{\rangeval{6.78}} & \makebox[20.25pt][r]{\textbf{1.68}}\hspace{1.4pt}\makebox[18.125pt][l]{\rangeval{0.07}} & -- & --\\[1.4pt]
\textsc{Materialized} & -- & -- & -- & -- & -- & -- & \makebox[29.25pt][r]{1976.17}\hspace{1.4pt}\makebox[25.625pt][l]{\rangeval{123.16}} & \makebox[20.25pt][r]{\textbf{32.70}}\hspace{1.4pt}\makebox[18.125pt][l]{\rangeval{0.43}}\\[1.4pt]
\bottomrule
\end{tabular}\unskip
\end{table*}

The artifact provides the envelope statistics, absolute timings, and memory measurements underlying the comparisons below, including results not plotted in the figures.

\noindent\textbf{Real-Data Queries.}
On the original STATS queries, our sampler improves PRODUCT sampling
latency more consistently than total method cost (Fig.~\ref{fig:stats}).
It samples faster than the comparison baseline in 17 of 18 PRODUCT cases,
but reduces $B$$+$$kq$ in 9 of 18 under the scheduled call counts.
The real PRODUCT envelope masses help explain the sampling benefit:
\textsc{Global} uses about $2.0$--$4.6$ times $\mu$ and \textsc{Root}
uses $1.2$--$1.7$ times $\mu$, whereas our proposal-rate multiplier $C$
is at most about $1.08$. Fewer proposals reduce the work spent generating
and rejecting candidates. Real-feature weights yield a sampling benefit
in 8 of 9 cases, with Q064 at $\mu$$=$$10{,}000$ the exception
(sampling-time ratio about $1.16$).
Across all 36 STATS cases, construction takes
$1.08$--$1.26$ times the fastest baseline's time,
so faster calls must first recover that extra cost.
Across all probability modes, sampling time is lower in 28 of 36 cases
and total cost in 10 of 36.

On NEWpath4, real PRODUCT sampling-time ratios are about $0.50$, $0.49$,
and $0.87$ at increasing expected sample sizes. Total method cost is lower
at $\mu=100$ and $10{,}000$, with ratios $0.79$ and $0.88$. At $\mu=1$,
1000 calls leave the construction cost higher than the savings in sampling.
Under uniform probabilities, the envelopes are tight and our sampler's
median sampling-time ratio to \textsc{Rank} is $0.999$ across 9 cases.
Our allocated index memory is smaller than \textsc{Score}, \textsc{Global},
and \textsc{Root} in all 36 cases. With applicable \textsc{Rank} and
\textsc{Scan} indexes also considered, the count falls to 21 of 36.

JOB1a combines a large filtered input with just 142 join results and a
\textsc{Root} envelope mass equal to $\mu$. Our sampler has lower sampling
time in 4 of 10 cases, but higher total method cost in all 10.
Our index occupies about 854\,MiB, compared with
$0.0072$\,MiB for \textsc{Scan}. The peak RSS ratio is much closer, about
$1.07$, because \textsc{Scan} also holds the common input and needs
construction workspace. Table~\ref{tab:method-results} reports absolute
construction and sampling times for this query, NEWpath4, and a triangular
case, with the paired materialization measurements in the last two columns.

\begin{figure*}[!t]
\centering
\includegraphics[width=0.98\textwidth]{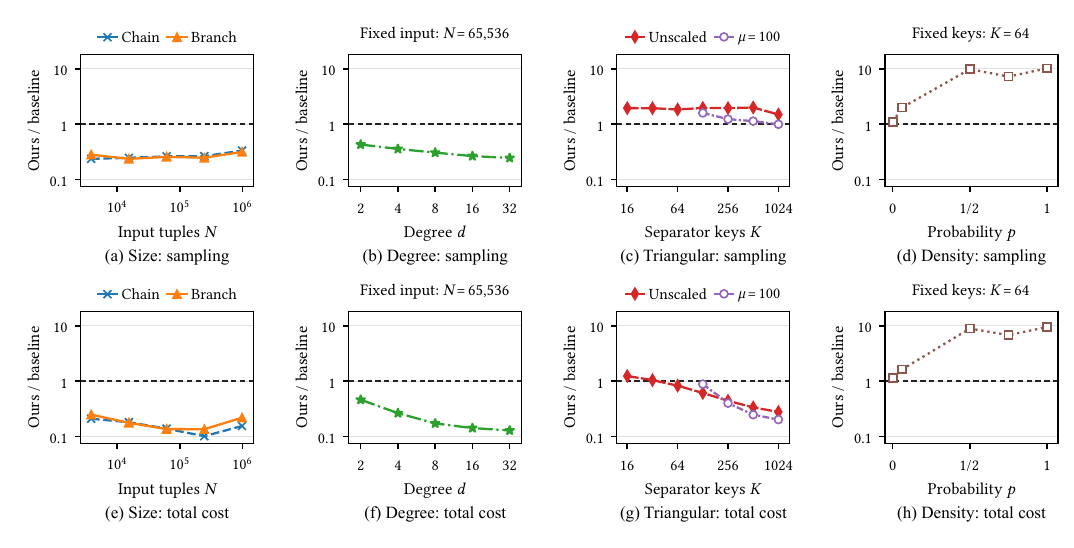}
\vspace{-2mm}
\caption{All 31 controlled cases. Panels (a)--(d) show sampling time and (e)--(h) total cost. Points are ratios of four-process medians against the fastest applicable baseline with complete measurements for each metric. Ratio axes are logarithmic, and black dashed lines mark a ratio of one. Four \textsc{Global} cases exceeded the process time limit in every repeat, so their unmeasured times are excluded from baseline selection.}
\label{fig:controlled}
\end{figure*}

\noindent\textbf{Controlled Workloads.}
At fixed input size and expected sample size $\mu=100$, our sampling time
stays between about $0.05$ and $0.07$\,ms while $J$ grows from roughly
131 thousand to 537 million in the degree family.
A much larger implicit result therefore
need not translate into more work per sampling call. Our sampler is
faster in both sampling and total method cost in all 15 size and degree cases
(Fig.~\ref{fig:controlled}).

The triangular cases separate sampling speed from construction savings.
\textsc{Score} samples faster in 10 of 11 cases, while cheaper construction
lets our sampler reduce total method cost in 9 of the 11 triangular cases.
Loose \textsc{Global} and \textsc{Root} envelopes alone therefore do not
determine which sampler has the fastest calls. All 16 timeouts are the 4
\textsc{Global} repeats in 4 triangular cases. Only process-level timeout
status is available for them, so their construction and sampling times are
unmeasured.

Dense samples reduce the benefit of selective sampling because all methods
must write many results. Baselines have lower sampling time and total cost
in all 5 dense cases. At $p=1$, \textsc{Scan} is about $10.2$ times faster
per call, illustrating the importance of implementation constants when
the entire join must be returned.

\noindent\textbf{Input and Join Size.}
NEWbranch5 and JOB17a contrast a large implicit join with a larger input
whose join can be stored explicitly. Their results show why both sizes
and the tightness of probability bounds matter.
On NEWbranch5, our sampler has lower sampling time in all 3 real
PRODUCT cases. Ratios to the comparison baseline are about $0.38$, $0.38$,
and $0.71$ at increasing expected sample sizes. Total method cost ratios
are $1.12$, $0.75$, and $0.73$, so the smallest sample again offers too little
sampling work to repay construction within the scheduled calls. Even the
best root has envelope mass about $2.23\mu$, leaving substantial rejection work
for \textsc{Root}. At $\mu=100$, our sampler uses about 31\,MiB of allocated
index memory and 136\,MiB of peak RSS. Across all 5 probability settings,
it samples faster in every NEWbranch5 case and has lower total method cost
in 2. Table~\ref{tab:extension-results} reports absolute
times at $\mu=100$. The artifact includes all five configurations per query and their memory measurements. 

\begin{table*}[t]
\centering
\caption{Extension batch, real PRODUCT, $\mu=100$. Construction time $B$ (ms) and sampling time $q$ ($\mu$s) use the four-process median (full observed range) notation of Table~\ref{tab:method-results}. Bold marks the lowest median per metric and query. NA: exceeds the output budget. MEM 4/4: all four processes exhausted 32\,GiB, leaving $B$ and $q$ unmeasured. \textsc{Rank} applies only to uniform probabilities.}
\label{tab:extension-results}
\fontsize{9}{10.8}\selectfont
\setlength{\tabcolsep}{8pt}
\begin{tabular*}{0.99\textwidth}{@{\extracolsep{\fill}}lcccc@{}}
\toprule
\textbf{Method} & \multicolumn{2}{c}{\shortstack{\textbf{NEWbranch5}}} & \multicolumn{2}{c}{\shortstack{\textbf{JOB17a}}}\\
\cmidrule(lr){2-3}\cmidrule(lr){4-5}
 & $B$ (ms) & $q$ ($\mu$s) & $B$ (ms) & $q$ ($\mu$s)\\\midrule
\textsc{Ours} & \makebox[24.75pt][r]{88.01}\hspace{1.4pt}\makebox[18.125pt][l]{\rangeval{0.85}} & \makebox[24.75pt][r]{\textbf{78.48}}\hspace{1.4pt}\makebox[18.125pt][l]{\rangeval{1.49}} & \makebox[33.75pt][r]{6803.89}\hspace{1.4pt}\makebox[25.625pt][l]{\rangeval{112.52}} & \makebox[29.25pt][r]{180.62}\hspace{1.4pt}\makebox[21.875pt][l]{\rangeval{2.09}}\\[1.4pt]
\textsc{Score} & \makebox[24.75pt][r]{831.56}\hspace{1.4pt}\makebox[18.125pt][l]{\rangeval{4.40}} & \makebox[24.75pt][r]{235.56}\hspace{1.4pt}\makebox[18.125pt][l]{\rangeval{3.11}} & MEM 4/4 & MEM 4/4\\[1.4pt]
\textsc{Global} & \makebox[24.75pt][r]{\textbf{73.79}}\hspace{1.4pt}\makebox[18.125pt][l]{\rangeval{0.36}} & \makebox[24.75pt][r]{346.79}\hspace{1.4pt}\makebox[18.125pt][l]{\rangeval{3.46}} & \makebox[33.75pt][r]{\textbf{5704.67}}\hspace{1.4pt}\makebox[25.625pt][l]{\rangeval{19.53}} & \makebox[29.25pt][r]{2483.72}\hspace{1.4pt}\makebox[21.875pt][l]{\rangeval{35.92}}\\[1.4pt]
\textsc{Root} & \makebox[24.75pt][r]{85.87}\hspace{1.4pt}\makebox[18.125pt][l]{\rangeval{0.45}} & \makebox[24.75pt][r]{207.31}\hspace{1.4pt}\makebox[18.125pt][l]{\rangeval{2.89}} & \makebox[33.75pt][r]{6144.73}\hspace{1.4pt}\makebox[25.625pt][l]{\rangeval{43.89}} & \makebox[29.25pt][r]{203.29}\hspace{1.4pt}\makebox[21.875pt][l]{\rangeval{3.31}}\\[1.4pt]
\textsc{Scan} & NA & NA & \makebox[33.75pt][r]{17126.39}\hspace{1.4pt}\makebox[25.625pt][l]{\rangeval{272.69}} & \makebox[29.25pt][r]{1958.85}\hspace{1.4pt}\makebox[21.875pt][l]{\rangeval{8.04}}\\[1.4pt]
\textsc{Materialized} & NA & NA & \makebox[33.75pt][r]{16003.24}\hspace{1.4pt}\makebox[25.625pt][l]{\rangeval{72.37}} & \makebox[29.25pt][r]{\textbf{19.48}}\hspace{1.4pt}\makebox[21.875pt][l]{\rangeval{0.50}}\\[1.4pt]
\bottomrule
\end{tabular*}\unskip
\end{table*}

JOB17a has a much larger input but a join small enough to store explicitly.
A cast\_info tuple determines its title, making the PRODUCT probability
constant within each chosen root block and \textsc{Root}'s envelope mass
equal to $\mu$. \textsc{Materialized} samples fastest in all 3 PRODUCT
cases. Our sampler takes about $9.5$, $9.3$, and $16.7$ times as long.
\textsc{Global} has the lowest total method cost at $\mu$$=$$1$, and
\textsc{Root} at the two larger sample sizes. Our cost ratios to these
baselines are $1.19$, $1.11$, and $1.21$, despite cheaper construction than
\textsc{Materialized}. At $\mu$$=$$10{,}000$, the maximum probability is
$0.606$, activating the heavy-witness branch. Our sampler, \textsc{Root},
and \textsc{Materialized} respectively take about $22.7$, $13.6$, and
$1.4$\,ms per call.

At $\mu=100$, their indexes occupy about $2418$, $4886$, and $12$\,MiB,
and peak RSS is about $11{,}100$, $13{,}100$, and $12{,}100$\,MiB.
The common input and construction workspace dominate the latter metric.
All 12 \textsc{Score} processes in the 3 PRODUCT cases exhaust
the 32\,GiB process memory limit, leaving $B$ and $q$ unmeasured, while
every uniform and one-relation repeat completes. The same
sampling-time and total-cost disadvantage for our sampler holds in the
uniform and one-relation settings, covering all 5 JOB17a cases.

\subsection{Materialization and Repeated Sampling}

\begin{figure*}[!t]
\centering
\includegraphics[width=0.98\textwidth]{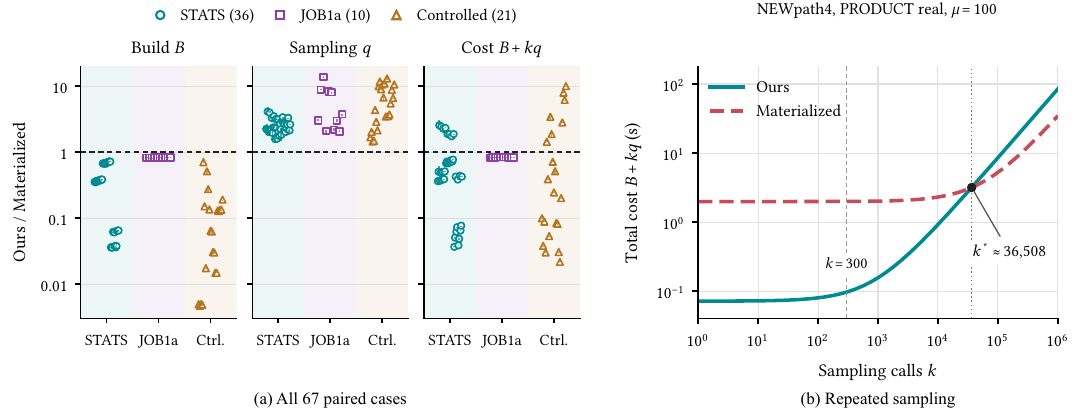}
\vspace{-1mm}
\caption{Paired comparison with \textsc{Materialized}. (a) All 67 cases: markers show medians of four paired ratios, and vertical segments show their observed ranges. (b) NEWpath4 with real PRODUCT weights: curves evaluate $B+kq$ using measured construction and sampling-time medians. The intersection estimates the break-even call count. Both axes in (b) are logarithmic.}
\label{fig:materialized}
\end{figure*}

How many sampling calls are needed to repay the construction cost of
an explicit subset index? In all 67 paired cases, our index builds faster
and \textsc{Materialized} has lower sampling time. Under the scheduled
call counts, our sampler has lower total method cost in 52 cases:
28 of 36 STATS, all 10 JOB1a, and 14 of 21 controlled cases
(Fig.~\ref{fig:materialized}). Allocated index memory is lower in
44 cases and RSS in 53.

To estimate when the advantage changes, let subscripts $\mathrm{O}$ and
$\mathrm{M}$ denote our sampler and
\textsc{Materialized}. Because $B_{\mathrm{O}}<B_{\mathrm{M}}$ and
$q_{\mathrm{O}}>q_{\mathrm{M}}$, the median components favor our sampler
for $k<k^*=(B_{\mathrm{M}}-B_{\mathrm{O}})/(q_{\mathrm{O}}-q_{\mathrm{M}})$.
The crossover $k^*$ is calculated from measured construction and
sampling-time medians. For NEWpath4 with real PRODUCT weights, it is
about 36,500 calls at $\mu=100$ and 300 calls at $\mu=10{,}000$.
Fewer calls favor our cheaper construction, while more calls allow
\textsc{Materialized} to recover its initial cost.

At $\mu=100$, the paired measurements in Table~\ref{tab:method-results}
imply about 97\,ms for our sampler and 1.99\,s for \textsc{Materialized}
over 300 calls. The indexes occupy about 22.2 and 816.6\,MiB, and peak
RSS is 111.8 and 970.4\,MiB. At $\mu=10{,}000$, total method costs over
100 calls are about 952\,ms and 2.18\,s.
Overall, 64 paired cases have an interval of positive integer call counts
in which our sampler is cheaper. In the three dense cases with
$p=1/2,3/4,1$, the estimated crossovers are only $0.237$, $0.221$, and
$0.102$ calls, respectively, so \textsc{Materialized} is already cheaper
at the first call.

The paired study compares only these two methods. JOB1a illustrates why
the comparator matters: our sampler beats \textsc{Materialized} in total
cost in all 10 paired cases, but other baselines have lower cost in all
10 original cases. 

Per-case memory measurements and crossover estimates are provided in the artifact.

\subsection{Numerical Implementation and Validation}
We check how the floating-point prototype handles numerical boundaries
and whether its output agrees with the prescribed subset distribution
on the tested instances. The checks cover full subsets and independence,
as well as output counts in the performance experiments.

For positive $p$, our sampler evaluates $-\operatorname{log1p}(-p)/p$,
using value one at zero.\footnote{The function $\operatorname{log1p}(x)$
evaluates $\log(1+x)$ accurately near zero.}
With machine precision $\eta=2^{-52}$, it inflates
$C$ by $1+64m\eta$ and rounds upward once. Heavy search uses cutoff
$(1/2)/(1+128m^2\eta)$. Terminal heavy classification and light rejection
use the same preorder product and strict $p>1/2$ comparison, keeping
the two paths consistent near the threshold. Endpoint coins
return false at $p=0$ and true at $p=1$. The implementation rejects nonfinite
total weights or rates and invalid thinning probabilities. Underflow has
no general recovery mechanism.
The prototype allocates its $m$-entry scratch eagerly. The proof uses
lazy initialization to obtain the displayed dependence on $m$, and both
satisfy the stated bound for a fixed query $Q$.

Small enumerable joins let us compare the distribution of the entire
returned subset with the product-Bernoulli law. We use total variation
distance, half the sum of absolute differences between the two
subset probabilities over all 
possible subsets.
A three-relation, 13-row chain
with five witnesses of nominal probabilities $(1,0.5,0.72,0.2,0)$, the first
four sharing a root row, tests independence despite shared input tuples.
Over 250,000 calls, the empirical distribution on
all 32 subsets has total variation distance about $0.0029$ from the
product-Bernoulli law. The suite also checks marginals, within-call pairs,
and each witness's frequency in adjacent call pairs against $p^2$.

To test the heavy/light boundary, a second
three-relation chain has 11 rows and nine witnesses sharing their first 2
rows. Last-row weights are perturbed by up to four units in the last place
around a product of $1/2$. Over 150,000 calls, the empirical law on all 512
subsets has total variation distance about $0.024$.
Deterministic comparisons with an enumerated join verify the heavy
set and its probabilities in 14,502 cases. We also compare SQL and join-tree
counts and run sanitizer checks.

In the performance experiments, output counts provide a broader check
across methods and cases. Let $T$ be the number of measured calls across
successful repeats and $Y$ their total output count. We compute the
standardized count residual
\[
R_{\mathrm{count}}=\frac{Y-T\mu}{\sqrt{T\bigl(\mu-\sum_t p(t)^2\bigr)}},
\]
using the independent-Bernoulli count variance. Zero-variance cases are
checked against their deterministic count. The largest absolute residuals
are about $2.55$ for original STATS, $2.53$ for controlled cases, $3.60$ for
JOB1a, and $2.82$ for the paired study.
The extension maximum is $2.88$. We use these maxima as descriptive checks
of sample counts across measured runs. The artifact includes the full numerical-check outputs and the measurements underlying the count residuals.

\section{Related Work}
\label{sec:related}

\noindent\textbf{Join Sampling.}
Direct sampling from a join uses its structure and input frequencies to
select compatible tuples without enumerating the full result.
Chaudhuri et al.~\cite{chaudhuri-motwani-narasayya-1999} establish the role
of cross-relation frequency information, while Zhao
et al.~\cite{zhao-christensen-li-hu-yi-2018} examine preprocessing and
rejection tradeoffs for uniform multiway join sampling. For weighted
results, the multistage methods of Shekelyan
et al.~\cite{shekelyan-cormode-ma-shanghooshabad-triantafillou-2023} support
product-weighted proportional draws. Deng et al.~\cite{deng-lu-tao-2023}
further connect uniform join sampling to the hardness of combinatorial
output-sensitive join algorithms.
Related query samplers support a requested number of draws from spatial
range joins~\cite{amagata-2025-spatial} and interval
queries~\cite{amagata-2024-interval}.
We use proportional draws to generate a subset with prescribed
probabilities and mutually independent inclusion decisions.

Esmailpour et al.~\cite{esmailpour-hu-huang-sintos-2026} directly study
independent subset sampling over joins, covering PRODUCT, MIN, MAX, and
SUM probabilities and insertion-only maintenance. For static PRODUCT
probabilities on fixed full acyclic joins, our algorithm simultaneously
removes the logarithmic factors in preprocessing, index space, and
expected sampling time under the exact-real model of
Section~\ref{sec:model}. Their score buckets group output probabilities
and support direct access within each bucket. Our shared index uses total
weights and maximum completion weights for separator groups to support
both proportional draws and high-probability witness enumeration in
linear space.
Bekkers et al.~\cite{bekkers-neven-pantelis-vansummeren-2026} address
Poisson sampling over acyclic joins when one relation supplies the
probability attribute. Our PRODUCT formulation also permits weights across
multiple relations, with one-relation experiments isolating the
specialized setting.
Huang et al.~\cite{huang-tao-wang-2026-selections} study acyclic join
sampling under equality selections supplied at query time, whereas our
index is built for fixed filtered inputs.

\noindent\textbf{Join Trees and Ranked Enumeration.}
Our index follows the join-tree organization of acyclic
evaluation~\cite{yannakakis-1981}. Its high-probability path needs to
enumerate all witnesses above a probability threshold in arbitrary order.
Ranked enumeration over full conjunctive queries uses dynamic programming and best
completions~\cite{tziavelis-ajwani-gatterbauer-riedewald-yang-2020}, and
extends factorization and ranking beyond
equijoins~\cite{tziavelis-gatterbauer-riedewald-2021}. We specialize this
maximum-completion approach to threshold enumeration.
Coupling this traversal with proportional draws handles high probabilities
directly and bounds proposal work for the remaining witnesses.

\noindent\textbf{Subset Sampling and Computational Models.}
For explicitly stored probabilities, Bringmann and
Panagiotou~\cite{bringmann-panagiotou-2017} achieve optimal expected
output-sensitive subset sampling. Applying such a structure to an implicit
join first requires constructing and storing its witnesses.
Extensions of explicit subset sampling support dynamic
sets~\cite{yi-wang-wei-2023,bhattacharya-kiss-sidford-wajc-2024},
probabilities derived from normalized dynamic
weights~\cite{huang-wang-2025-dips}, and external-memory or range
queries~\cite{huang-wang-2023-subset-extensions}. Parallel algorithms also
support proportional and subset
interfaces~\cite{huebschle-schneider-sanders-2022}. Our
\textsc{Materialized} baseline measures the construction and repeated-sampling
tradeoff of applying an explicit subset index to a join.

Exact random generation depends on the operations a computational model
permits and how their costs are charged. On bounded-word machines,
Bringmann and Friedrich~\cite{bringmann-friedrich-2013} develop exact
geometric and Bernoulli generation, while the parameterized subset sampler
of Gan et al.~\cite{gan-umboh-wang-wirth-zhang-2024} provides exact
Bernoulli and truncated-geometric primitives. Our theorem uses unit-cost
exact real arithmetic, logarithms, and continuous uniform randomness
(Section~\ref{sec:model}). The floating-point prototype implements
numerical approximations to that model.

\noindent\textbf{Poisson Constructions and Graph Generation.}
Our light-witness construction simulates arrivals with proportional
selection~\cite{gillespie-1976} and independent
thinning~\cite{lewis-shedler-1979}. The Bernoulli--Poisson
link~\cite{zhou-2015} turns the resulting accepted counts into inclusion
decisions. Related graph generators use products of endpoint weights in
rank-one models~\cite{miller-hagberg-2011}, or structured products in Kronecker
models~\cite{moreno-pfeiffer-neville-kirshner-2014}.
Ramani et al.~\cite{ramani-eikmeier-gleich-2019} distinguish exact skipping
from ordinary edge draws followed by deduplication. More recently, the
edge-arrival construction of D'Angelo and
Michielan~\cite{dangelo-michielan-2026}, currently a preprint, combines
Poisson counts, alias proposals, splitting, and deduplication for
expected-degree graphs.

General acyclic joins introduce equality constraints that determine which
input tuple combinations are valid. Our shared index supports proportional
selection over these valid witnesses and locates high-probability
witnesses without materializing the join. Combining these operations with
the probability construction yields expected linear preprocessing, linear
index space, and output-linear expected sampling work in our model.

\noindent\textbf{Dynamic and Projected Queries.}
Maintaining samples under updates requires controlling the work caused by
changed input tuples. Stream sampling under arbitrary
updates~\cite{tao-lian-papadias-hadjieleftheriou-2007}, maintained join
synopses~\cite{zhao-li-liu-2020}, and insertion-only join
reservoirs~\cite{dai-hu-yi-2026} address this task. The cost of maintaining
query structures also depends on structural tractability
boundaries~\cite{berkholz-keppeler-schweikardt-2017} and heavy-light
decompositions based on input
degrees~\cite{abo-khamis-chmielewski-draghici-kara-olteanu-2026}.
We study static inputs and weights, allowing index summaries to be built
once and reused across independent sampling calls.
Projection raises a separate issue of output identity: multiple full
witnesses can represent one distinct projected answer.
Hu and Huang~\cite{hu-huang-2026} study sampling at this projected-answer
level. Our algorithm deduplicates full occurrence vectors and samples at
the witness level.

\section{Conclusion}
\label{sec:conclusion}

We presented an independent PRODUCT subset sampler for fixed full acyclic
joins with static weights. A shared join-tree index supports proportional
draws and high-probability witness enumeration, giving expected linear
preprocessing, linear index space, and output-linear expected sampling work.
Under the stated exact-real model, the time bounds are optimal:
preprocessing plus the first call matches input reading and expected output
writing, while subsequent calls match expected output writing up to constant
overhead. Experiments with our floating-point prototype show how
construction savings affect total cost over repeated calls.
In 52 of 67 paired configurations, our sampler has lower
construction-plus-sampling cost than indexed materialization under the
scheduled call counts, excluding common preparation and loading.
Indexed materialization answers individual calls faster, so our construction
savings can favor workloads with limited reuse. With enough reuse, faster
calls can repay the higher construction cost.

\typeout{ICDEMAINPAGES=\number\numexpr\value{page}-1\relax}


\begin{thebibliography}{10}
\providecommand{\url}[1]{#1}
\csname url@samestyle\endcsname
\providecommand{\newblock}{\relax}
\providecommand{\bibinfo}[2]{#2}
\providecommand{\BIBentrySTDinterwordspacing}{\spaceskip=0pt\relax}
\providecommand{\BIBentryALTinterwordstretchfactor}{4}
\providecommand{\BIBentryALTinterwordspacing}{\spaceskip=\fontdimen2\font plus
\BIBentryALTinterwordstretchfactor\fontdimen3\font minus
  \fontdimen4\font\relax}
\providecommand{\BIBforeignlanguage}[2]{{%
\expandafter\ifx\csname l@#1\endcsname\relax
\typeout{** WARNING: IEEEtran.bst: No hyphenation pattern has been}%
\typeout{** loaded for the language `#1'. Using the pattern for}%
\typeout{** the default language instead.}%
\else
\language=\csname l@#1\endcsname
\fi
#2}}
\providecommand{\BIBdecl}{\relax}
\BIBdecl

\bibitem{chaudhuri-motwani-narasayya-1999}
S.~Chaudhuri, R.~Motwani, and V.~Narasayya, ``On random sampling over joins,''
  in \emph{Proceedings of the ACM SIGMOD International Conference on Management
  of Data}, 1999, pp. 263--274.

\bibitem{zhao-christensen-li-hu-yi-2018}
Z.~Zhao, R.~Christensen, F.~Li, X.~Hu, and K.~Yi, ``Random sampling over joins
  revisited,'' in \emph{Proceedings of the ACM SIGMOD International Conference
  on Management of Data}, 2018, pp. 1525--1539.

\bibitem{shekelyan-cormode-ma-shanghooshabad-triantafillou-2023}
M.~Shekelyan, G.~Cormode, Q.~Ma, A.~Mohammadi~Shanghooshabad, and
  P.~Triantafillou, ``Streaming weighted sampling over join queries,'' in
  \emph{Proceedings of the 26th International Conference on Extending Database
  Technology (EDBT)}, 2023, pp. 298--310.

\bibitem{bringmann-panagiotou-2017}
K.~Bringmann and K.~Panagiotou, ``Efficient sampling methods for discrete
  distributions,'' \emph{Algorithmica}, vol.~79, no.~2, pp. 484--508, 2017.

\bibitem{esmailpour-hu-huang-sintos-2026}
A.~Esmailpour, X.~Hu, J.~Huang, and S.~Sintos, ``Subset sampling over joins,''
  \emph{Proceedings of the ACM on Management of Data}, vol.~4, no.~2, 2026,
  art. 117.

\bibitem{bekkers-neven-pantelis-vansummeren-2026}
L.~Bekkers, F.~Neven, L.~Pantelis, and S.~Vansummeren, ``{Poisson} sampling
  over acyclic joins,'' \emph{Proceedings of the ACM on Management of Data},
  vol.~4, no.~3, 2026, art. 224.

\bibitem{yannakakis-1981}
M.~Yannakakis, ``Algorithms for acyclic database schemes,'' in
  \emph{Proceedings of the 7th International Conference on Very Large Data
  Bases (VLDB)}, 1981, pp. 82--94.

\bibitem{walker-1977}
A.~J. Walker, ``An efficient method for generating discrete random variables
  with general distributions,'' \emph{ACM Transactions on Mathematical
  Software}, vol.~3, no.~3, pp. 253--256, 1977.

\bibitem{vose-1991}
M.~D. Vose, ``A linear algorithm for generating random numbers with a given
  distribution,'' \emph{IEEE Transactions on Software Engineering}, vol.~17,
  no.~9, pp. 972--975, 1991.

\bibitem{gillespie-1976}
D.~T. Gillespie, ``A general method for numerically simulating the stochastic
  time evolution of coupled chemical reactions,'' \emph{Journal of
  Computational Physics}, vol.~22, no.~4, pp. 403--434, 1976.

\bibitem{lewis-shedler-1979}
P.~A.~W. Lewis and G.~S. Shedler, ``Simulation of nonhomogeneous {Poisson}
  processes by thinning,'' \emph{Naval Research Logistics Quarterly}, vol.~26,
  no.~3, pp. 403--413, 1979.

\bibitem{tziavelis-ajwani-gatterbauer-riedewald-yang-2020}
N.~Tziavelis, D.~Ajwani, W.~Gatterbauer, M.~Riedewald, and X.~Yang, ``Optimal
  algorithms for ranked enumeration of answers to full conjunctive queries,''
  \emph{Proceedings of the VLDB Endowment}, vol.~13, no.~9, pp. 1582--1597,
  2020.

\bibitem{zhou-2015}
M.~Zhou, ``Infinite edge partition models for overlapping community detection
  and link prediction,'' in \emph{Proceedings of the 18th International
  Conference on Artificial Intelligence and Statistics (AISTATS)}, ser.
  Proceedings of Machine Learning Research, vol.~38, 2015, pp. 1135--1143.

\bibitem{ieee-754-2019}
\BIBentryALTinterwordspacing
\emph{{IEEE} Standard for Floating-Point Arithmetic}, IEEE Std. 754-2019, 2019.
  [Online]. Available: \url{https://standards.ieee.org/ieee/754/6210/}
\BIBentrySTDinterwordspacing

\bibitem{join-subset-sampling}
\BIBentryALTinterwordspacing
{ICDE2027optimal}, ``Join subset sampling,'' GitHub repository, 2026. [Online].
  Available: \url{https://github.com/ICDE2027optimal/join-subset-sampling}
\BIBentrySTDinterwordspacing

\bibitem{han-et-al-2022-stats-ceb}
Y.~Han, Z.~Wu, P.~Wu, R.~Zhu, J.~Yang, L.~W. Tan, K.~Zeng, G.~Cong, Y.~Qin,
  A.~Pfadler, Z.~Qian, J.~Zhou, J.~Li, and B.~Cui, ``Cardinality estimation in
  {DBMS}: A comprehensive benchmark evaluation,'' \emph{Proceedings of the VLDB
  Endowment}, vol.~15, no.~4, pp. 752--765, 2022.

\bibitem{leis-et-al-2015-job}
V.~Leis, A.~Gubichev, A.~Mirchev, P.~Boncz, A.~Kemper, and T.~Neumann, ``How
  good are query optimizers, really?'' \emph{Proceedings of the VLDB
  Endowment}, vol.~9, no.~3, pp. 204--215, 2015.

\bibitem{deng-lu-tao-2023}
S.~Deng, S.~Lu, and Y.~Tao, ``On join sampling and the hardness of
  combinatorial output-sensitive join algorithms,'' in \emph{Proceedings of the
  42nd ACM SIGMOD-SIGACT-SIGAI Symposium on Principles of Database Systems
  (PODS)}, 2023, pp. 99--111.

\bibitem{amagata-2025-spatial}
D.~Amagata, ``Random sampling over spatial range joins,'' in \emph{IEEE 41st
  International Conference on Data Engineering (ICDE)}.\hskip 1em plus 0.5em
  minus 0.4em\relax IEEE, 2025, pp. 2080--2093.

\bibitem{amagata-2024-interval}
D.~Amagata, ``Independent range sampling on interval data,'' in \emph{2024 IEEE
  40th International Conference on Data Engineering (ICDE)}.\hskip 1em plus
  0.5em minus 0.4em\relax IEEE, 2024, pp. 449--461.

\bibitem{huang-tao-wang-2026-selections}
J.~Huang, Y.~Tao, and S.~Wang, ``Acyclic join sampling under selections:
  Dichotomy, union sampling, and enumeration,'' in \emph{29th International
  Conference on Database Theory (ICDT)}, ser. Leibniz International Proceedings
  in Informatics (LIPIcs), vol. 365, 2026, pp. 9:1--9:20.

\bibitem{tziavelis-gatterbauer-riedewald-2021}
N.~Tziavelis, W.~Gatterbauer, and M.~Riedewald, ``Beyond equi-joins: Ranking,
  enumeration and factorization,'' \emph{Proceedings of the VLDB Endowment},
  vol.~14, no.~11, pp. 2599--2612, 2021.

\bibitem{yi-wang-wei-2023}
L.~Yi, H.~Wang, and Z.~Wei, ``Optimal dynamic subset sampling: Theory and
  applications,'' in \emph{Proceedings of the 29th ACM SIGKDD Conference on
  Knowledge Discovery and Data Mining}, 2023, pp. 3116--3127.

\bibitem{bhattacharya-kiss-sidford-wajc-2024}
S.~Bhattacharya, P.~Kiss, A.~Sidford, and D.~Wajc, ``Near-optimal dynamic
  rounding of fractional matchings in bipartite graphs,'' in \emph{Proceedings
  of the 56th Annual ACM Symposium on Theory of Computing (STOC)}, 2024, pp.
  59--70.

\bibitem{huang-wang-2025-dips}
J.~Huang and S.~Wang, ``{DIPS}: Optimal dynamic index for {Poisson} {$\pi$}ps
  sampling,'' in \emph{Proceedings of the 31st ACM SIGKDD Conference on
  Knowledge Discovery and Data Mining}, 2025, pp. 520--531.

\bibitem{huang-wang-2023-subset-extensions}
\BIBentryALTinterwordspacing
J.~Huang and S.~Wang, ``Subset sampling and its extensions,'' arXiv preprint
  arXiv:2307.11585, 2023. [Online]. Available:
  \url{https://arxiv.org/abs/2307.11585}
\BIBentrySTDinterwordspacing

\bibitem{huebschle-schneider-sanders-2022}
L.~H{\"u}bschle-Schneider and P.~Sanders, ``Parallel weighted random
  sampling,'' \emph{ACM Transactions on Mathematical Software}, vol.~48, no.~3,
  2022, art. 29.

\bibitem{bringmann-friedrich-2013}
K.~Bringmann and T.~Friedrich, ``Exact and efficient generation of geometric
  random variates and random graphs,'' in \emph{Automata, Languages, and
  Programming (ICALP)}, ser. Lecture Notes in Computer Science, vol. 7965,
  2013, pp. 267--278.

\bibitem{gan-umboh-wang-wirth-zhang-2024}
J.~Gan, S.~W. Umboh, H.~Wang, A.~Wirth, and Z.~Zhang, ``Optimal dynamic
  parameterized subset sampling,'' \emph{Proceedings of the ACM on Management
  of Data}, vol.~2, no.~5, 2024, art. 209.

\bibitem{miller-hagberg-2011}
J.~C. Miller and A.~Hagberg, ``Efficient generation of networks with given
  expected degrees,'' in \emph{Algorithms and Models for the Web Graph (WAW)},
  ser. Lecture Notes in Computer Science, vol. 6732, 2011, pp. 115--126.

\bibitem{moreno-pfeiffer-neville-kirshner-2014}
S.~Moreno, J.~J. Pfeiffer, III, J.~Neville, and S.~Kirshner, ``A scalable
  method for exact sampling from {Kronecker} family models,'' in \emph{IEEE
  International Conference on Data Mining (ICDM)}, 2014, pp. 440--449.

\bibitem{ramani-eikmeier-gleich-2019}
A.~S. Ramani, N.~Eikmeier, and D.~F. Gleich, ``Coin-flipping, ball-dropping,
  and grass-hopping for generating random graphs from matrices of edge
  probabilities,'' \emph{SIAM Review}, vol.~61, no.~3, pp. 549--595, 2019.

\bibitem{dangelo-michielan-2026}
\BIBentryALTinterwordspacing
G.~D'Angelo and R.~Michielan, ``Efficient generation of expected-degree graphs
  via edge-arrivals,'' arXiv preprint arXiv:2604.21504, 2026. [Online].
  Available: \url{https://arxiv.org/abs/2604.21504}
\BIBentrySTDinterwordspacing

\bibitem{tao-lian-papadias-hadjieleftheriou-2007}
Y.~Tao, X.~Lian, D.~Papadias, and M.~Hadjieleftheriou, ``Random sampling for
  continuous streams with arbitrary updates,'' \emph{IEEE Transactions on
  Knowledge and Data Engineering}, vol.~19, no.~1, pp. 96--110, 2007.

\bibitem{zhao-li-liu-2020}
Z.~Zhao, F.~Li, and Y.~Liu, ``Efficient join synopsis maintenance for data
  warehouse,'' in \emph{Proceedings of the ACM SIGMOD International Conference
  on Management of Data}, 2020, pp. 2027--2042.

\bibitem{dai-hu-yi-2026}
B.~Dai, X.~Hu, and K.~Yi, ``Reservoir sampling over joins,'' \emph{ACM
  Transactions on Database Systems}, vol.~51, no.~2, 2026, art. 9.

\bibitem{berkholz-keppeler-schweikardt-2017}
C.~Berkholz, J.~Keppeler, and N.~Schweikardt, ``Answering conjunctive queries
  under updates,'' in \emph{Proceedings of the 36th ACM SIGMOD-SIGACT-SIGAI
  Symposium on Principles of Database Systems (PODS)}, 2017, pp. 303--318.

\bibitem{abo-khamis-chmielewski-draghici-kara-olteanu-2026}
M.~Abo~Khamis, E.~Chmielewski, A.~Draghici, A.~Kara, and D.~Olteanu,
  ``Maintaining queries under updates using heavy-light partitioning of the
  input relations,'' \emph{Proceedings of the ACM on Management of Data},
  vol.~4, no.~2, 2026, art. 109.

\bibitem{hu-huang-2026}
X.~Hu and J.~Huang, ``Towards output-optimal uniform sampling and approximate
  counting for join-project queries,'' \emph{Proceedings of the ACM on
  Management of Data}, vol.~4, no.~2, 2026, art. 120.

\end{thebibliography}
\end{document}